\documentclass[11pt]{article}
\usepackage{graphicx}
\def\slash#1{\setbox0=\hbox{$#1$}#1\hskip-\wd0\hbox to\wd0{\hss\sl/\/\hss}}
\begin{document}
\title{ The Proton Electromagnetic Form Factor $F_2$ and Quark Orbital Angular
Momentum}

\author{Pankaj Jain$^1$ and John P. Ralston$^2$\\
$^1$ Physics Department, IIT, Kanpur, India 208016\\
$^2$Department of Physics \& Astronomy\\
University of Kansas, Lawrence, KS 66045} 

\maketitle

\begin{abstract}We analyze the
proton electromagnetic form factor ratio
$R(Q^{2})=QF_2(Q^{2})/F_1(Q^{2})$ as a function of momentum transfer
$Q^{2}$ within perturbative QCD. We find that the prediction for
$R(Q^{2})$ at large momentum transfer $Q$ depends on the exclusive
quark wave functions, which are unknown.  For a wide range of wave
functions we find that $ QF_2/F_1 \sim\ const$ at large momentum
transfer, in agreement with recent JLAB data.  
\end{abstract}

The recent JLAB measurement of the Proton Electromagnetic Form Factor ratio
\cite{jones,Gayou,Perdrisat} shows the puzzling behavior
$R(Q^2) = QF_2(Q^{2})/F_1(Q^{2}) \sim const. $ at large $Q$. The form factors 
$F_1$ and $F_2$
are defined by the standard relation,
\begin{equation}
<p', \, s'| J^{\mu}|p, \, s>= \bar N(P^\prime,s^\prime)
\left(\gamma^\mu F_1(Q^2) + i \sigma^{\mu\nu}q_\nu{F_2(Q^2)\over 2M}
\right)N(P,s)
\end{equation}
where $q=p^\prime - p$, $Q^2 = -q^2$, $J^\mu$ is the electromagnetic
current of the proton and $s$ and $s^\prime$ refer to the spins
of the initial and final proton. $F_1$ and $F_2$ are refered to as the Dirac and
Pauli form factors respectively.

The amplitude $i\bar N(p', \, s')F_{2}(Q^{2}) \sigma^{\mu\nu}q_{\nu}
N(p,s)$ represents the amplitude for {\it chirality} of the proton to
flip under momentum transfer $Q$.  This flip can occur due to flip in
the chirality of the proton's constituents.  However at large momentum
transfer $Q>>m_q$, where $m_q  \sim $ few MeV is the mass of the
quarks, the amplitude for a quark chirality flip is negligible.  Alternatively
the chirality of the proton can flip due to quark orbital angular
momentum \cite{BNL93,Buniy,Bologna,jain03}.  General principles
allow the existence of substantial quark orbital angular momentum
($OAM$), but a body of belief accumulated from the non-relativistic
quark model ($NRQM$) has not favored this possibility.  Note that the
3-dimensional $OAM$ of non-relativistic physics has awkward
Lorentz transformation properties, and we are concerned with
$OAM$ of an $SO(2)$ subgroup of rotations preserving particle momenta.
Here we analyze the non-zero quark $OAM$ contribution to $F_2$ in order to
determine pQCD predictions for the scaling behavior of the ratio
$R(Q)$.

The large $Q^2$ behavior of form factors is often discussed within the
Brodsky-Lepage\cite{BL80} factorization scheme.  In this scheme
the dominant contributions to amplitudes is assumed to come
from the asymptotically short-distance ($asd$) region $b\sim 1/Q$,
where $b$ is the transverse separation between quarks.  
Non-zero $OAM$ is excluded in the first step, and should not
contribute at large $Q^{2}$.  The $asd$ formalism is tested by the
hadron helicity conservation ($HHC$) rule \cite{BL}
$\lambda_A+\lambda_B=\lambda_C+ \lambda_D$ for the reaction
$A+B\rightarrow C+D$, where $\lambda_i$ is the helicity of the
particle `i'.  Failures of the $asd$ approach to correctly predict
experimental results are well known.  These include many
observed violations of the helicity conservation rule and the
observation of oscillations in fixed-angle proton-proton elastic
scattering\cite{RP}.  A success of the formalism is the correct
prediction of the $Q^{2}$ scaling behavior of many exclusive
processes.  Yet scaling can also be obtained from the quark counting
model of Brodsky and Farrar and Matveev {\it et al} \cite{BFMatveev},
without assuming the details of the $asd$ formalism \cite{BL80}.  Thus
it is not possible to conclude whether or not quarks exist with non-zero
$OAM$ by appealing to $asd$ models.

\begin{figure}[]
\centering
\includegraphics[]{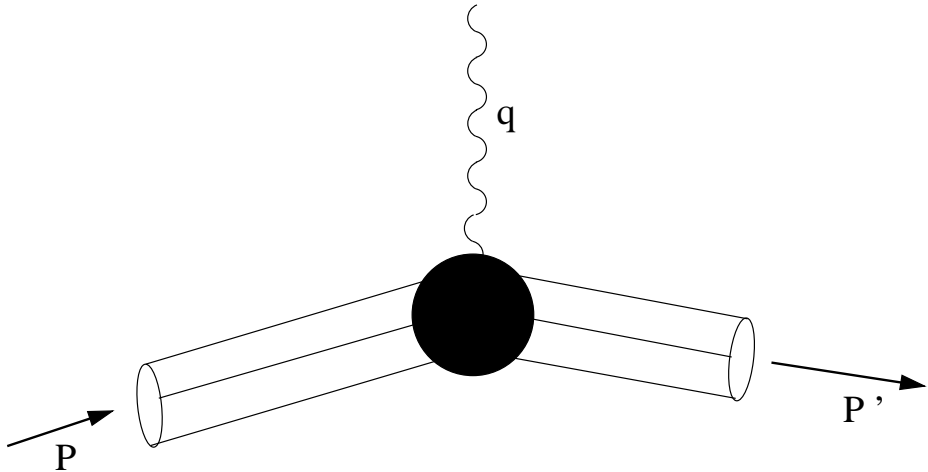}
\caption{Schematic illustration of the elastic scattering of a proton
from a virtual photon, $p(P) + \gamma^*(q) \rightarrow p(P^\prime)$.
	 }
\label{fig:gammap}
\end{figure}

The elastic scattering of a proton from a virtual photon is shown
schematically in Figure 1.  The three-quark contribution to the proton
form factor can be written in pQCD as
\begin{eqnarray}
\bar N(P^\prime,s^\prime)
&&\left(\gamma^\mu F_1(Q^2)+ i \sigma^{\mu\nu}q_\nu{F_2(Q^2)\over 2M}
\right)N(P,s)
= \int (d{\bf k_T})(dx)(d{\bf k_T^\prime})(dx^\prime)
\nonumber\\
&& \bar Y_{\alpha^\prime\beta^\prime
\gamma^\prime}(P^\prime,{\bf k_{Ti}^\prime},s^\prime)
\Gamma^\mu_{\alpha^\prime\beta^\prime\gamma^\prime\alpha\beta\gamma}(q,
{\bf k_{Ti}},{\bf k_{Ti}^\prime})
Y_{\alpha\beta\gamma}(P,{\bf k_{Ti}},s)\ .
\label{Eq:FormFactor}
\end{eqnarray}
We have factored the amplitude into products of a hard scattering kernel
$\Gamma^\mu$ and soft initial and final state wave functions
$Y$ and $Y^\prime$
respectively. The argument ${\bf k_{Ti}}$ of the initial wave function 
and the hard
scattering refers to the momenta of the three quarks, ${\bf k_{T1}}, 
{\bf k_{T2}}$
and ${\bf k_{T3}}$ with similar definition of the argument 
${\bf k_{Ti}^\prime}$ of
the final state wave function.  We use the ``brick-wall" frame for our
calculations.  The integration measures are given by,
$(d{\bf k_T}) = d^2k_{T1}d^2k_{T2}d^2k_{T3}\delta^2({\bf k_{T1}}+
{\bf k_{T2}}+{\bf k_{T3}})$, 
$(dx) = dx_1dx_2dx_3\delta(x_1+x_2+x_3)$. 
Note that Eq. \ref{Eq:FormFactor} postpones any assumptions
about the dominance of any particular integration region of ${\bf k_{T}}$
or $x$.  If care is not taken, limit interchange errors leading to 
the $asd$ results can
result.

To extract the contribution due to quark OAM, it is convenient to work
directly with the coordinates ${\bf b_i}$ conjugate to the transverse
quark momenta ${\bf k_{Ti}}$.  
We choose coordinates so that the third quark (down) lies at
the origin \cite{Li}, i.e. ${\bf b_3} = 0$.  The wave function $\tilde
Y_{\alpha\beta\gamma}$ can be decomposed as a sum of terms \cite{Li}:
\begin{eqnarray}
\tilde Y_{\alpha\beta\gamma}(P,{\bf b_i},s) &=& {f_N\over 8\sqrt{2} N_c}
({\cal C}^1_{\alpha\beta\gamma}V(P,{\bf b_i}) +
{\cal C}^2_{\alpha\beta\gamma}A(P,{\bf b_i}) \nonumber\\
&+&{\cal C}^3_{\alpha\beta\gamma}T(P,{\bf b_i}) +
{\cal C}^4_{\alpha\beta\gamma}X(P,{\bf b_i}) + ... )
\label{wavefn}
\end{eqnarray}
We are concerned with terms which lead by power counting in large
momenta.  Under a Lorentz transformation along the momentum axes the
variables $b$ are invariant.  We therefore keep leading wave 
functions whether or not a
power of $b$ occurs.  Each power of $b_{T} \rightarrow b_{x} \pm i b_{y}$ can
be further decomposed into combinations of quark $OAM$.  (At this
point, $asd$ would reject powers of $b$ and $OAM\neq 0$.)  The first 
few operators
${\cal C}$ are given by
\begin{eqnarray}
{\cal C}^1_{\alpha\beta\gamma} &=& (\slash P C)_{\alpha\beta}
(\gamma_5 N)_\gamma\nonumber\\
{\cal C}^2_{\alpha\beta\gamma} &=& (\slash P\gamma_5 C)_{\alpha\beta}
N_\gamma\nonumber\\
{\cal C}^3_{\alpha\beta\gamma} &=&- (\sigma_{\mu\nu} P^\nu C)_{\alpha\beta}
(\gamma^\mu\gamma_5 N)_\gamma\nonumber\\
{\cal C}^4_{\alpha\beta\gamma} &=& i(\slash P\gamma_5 C)_{\alpha\beta}
(\slash b_1  N)_\gamma
\label{DiracOperators}
\end{eqnarray}
Here $C$ is the charge-conjugation matrix.  Note the operator ${\cal 
C}^4$ which depends on $b_1$. Here 
$b_i$ are four vectors with
transverse components equal to ${\bf b_i}$ and all other components
zero. Similar operators exist for other transverse
coordinates.

The Dirac and Pauli form factors can now be obtained from the Eq.
\ref{Eq:FormFactor}.  $F_1$ can be schematically written as
\begin{equation}
F_1(Q)=\int
(d{\bf b})(d{\bf b^\prime}) (dx) (dx^\prime) {\psi^*(b, \, x)}
\tilde H(b_i,b_i^\prime\
x,\ x',\ Q)\psi(b, \, x').
\end{equation}
Here $\tilde H(b_i,b_i^\prime,\ x,\ x',\ Q)$ is the Fourier transform
of the hard scattering kernel after projecting out the Dirac bilinear
covariant $\bar N\gamma^\mu N$, and $\psi$ is a linear combination of
the wave functions $V,A$ and $T$ defined in Eq.  \ref{wavefn}.  A
detailed calculation of $F_1$ at leading order in perturbation series,
neglecting the transverse momenta in the Dirac propagators, is given
in Ref.  \cite{Li}.

We focus on the $b$ integrations to obtain $F_2$ in the limit of zero
quark masses.  Contributing are wave functions such as $X$ (Eq.
\ref{wavefn} ), as well as the transverse momenta 
in the Dirac propagators. The scaling behaviour can be determined
by considering the leading order hard scattering kernel. 
Since we are working in the impact parameter
space, we need to take the Fourier transform of this 
kernel. In this case the transverse momentum factors such as 
$(k_{T1})_i$, which occur in the Dirac propagators, turn
into derivatives $i\partial/\partial b_{1i} $. The remaining 
Fourier transform has a form similar to what is obtained for the form 
factor $F_1$. The dependence on the impact parameter in this kernel
arises through the Bessel functions, such as 
$K_0(\sqrt{x_1x_1^\prime} Q \tilde b_{12} )$, where $\tilde
b_{12}=|{\bf b_1-b_{2} }|$. Taking the derivative of this kernel with respect
to $b_{1i}$ gives factors of the form, 
$$  
{({\bf b_{1}}-{\bf b_{2}})_i\over \tilde b_{12}}\ Q
\sqrt{x_1x_1^\prime } K_0^\prime(\sqrt{x_1x_1^\prime Q^{2}\tilde 
b_{12}^{2}} )\ .
$$ 
Besides this transverse separation dependence, 
an additional power of $\bf b$ arises from the operator 
${\cal C}^4$ in the wave function. 

The scaling of $F_{2}(Q^{2}) $, and the form factor ratio, hinges on
scaling of the effective transverse separation $b$  at large $Q$.
Counting powers of $Q$, including one for the prefactor
$q^{\mu}\sigma_{\mu \nu}$, we find that if
$b\sim 1/Q$ at large $Q$, then $F_2/F_1 \sim 1/Q^2$ in this limit. This
result has recently been confirmed in Ref.  \cite{BJY} which
adopts the $asd$ formalism from the start.  Alternatively if $b\sim\
constant$ then $Q F_2/F_1\sim const$ and $F_2\sim 1/Q^5$ in the same
limit.  Such scaling was predicted in Ref.  \cite{Buniy} and is also
seen in a relativistic quark model calculation \cite{miller}.

We turn to how the dominant region of $b$ actually scales with $Q$. The simpler
case of the pion form factor\cite{LiSterman} is instructive.  The one 
gluon exchange
kernel in this case can be written as
\begin{equation}
\tilde H(b,
x,\ x',\ Q)  =  8 \pi^2 C_{F}\alpha_{s}
K_{0}(\sqrt{xx'Q^{2}b^{2}}).
\end{equation}
where $\alpha_s$ is the strong coupling
and $C_F$ is the color factor.
At large $Q$ the dominant contribution is obtained from the region
\begin{equation}
\sqrt{xx'}Q b <1.
\end{equation}
In order to reproduce the $asd$ assumptions, average values of $x\sim
1/2$ for pion and $x \sim 1/3$ for proton can be assigned, converting
$\sqrt{xx'}Q b <1 \rightarrow Q b <1$.  Such estimates are not the same thing
as actually computing the result!  Indeed one cannot rule out the
possibility that in the limit of very large Q, $b\sim constant$ and
$\sqrt{xx^\prime}\sim 1/Q$, or {\it any intermediate combinations of
integration regions}, given the complexity of the problem.  Elsewhere
\cite{jain03} we have shown that it is not possible, in principle, to
determine the scaling of a power of $b$ without specifying the
interplay of $x$ and $b$-dependence in the wave functions.  Indeed the
``rules of pQCD'' require one to objectively treat wave functions as
{\it unknown non-perturbative} objects, to be determined from data.
{\it Constructing} wave functions to make the $asd$ results come out is
possible, but we find such logic circular.

Suppose one uses a Gaussian model for the wave function $\Phi(b, \,
x, \, x^\prime)=b^{A}e^{-b^{2}/(2 a^{2})}\phi(x, x^\prime)$ to represent the
wave functions cutting off large $b$.  The factor of $b^{A}$ is the
phase-space to find $A$ quarks close together from naive
quark-counting.  We probe effects of quark $OAM$ by calculating the
moment $<b(Q)>_\pi$ defined by
  \begin{equation}
  <b(Q)>_\pi = {\int dxdx^\prime d^2 b b {\cal F}_\pi(Q, \, x, \, 
x^\prime, \, b)\over F_\pi(Q^2)}.
  \label{Eq:momentpi}
  \end{equation}
Then the pion form factor
  $F_\pi(Q^2)$ is given by
  \begin{equation}
  F_{\pi} (Q^2) = \int dxdx^\prime d^2 b  {\cal F}_{\pi} (Q, \, x,\, x^\prime,\,
  b)
  \end{equation}
and $ {\cal F}_{\pi} = \tilde H\Phi$.  We change variables for the
longitudinal fractions to $\xi=\sqrt{xx'}$, $\zeta=x/x'$ and 
parameterize $\phi(\xi, \zeta) \sim
\xi^{r+1}(1-\xi)^{r+1}\phi(\zeta),$ as $\xi \rightarrow 1$: the
$\zeta$ dependence can be left unspecified.  By substituting this into
Eq.  \ref{Eq:momentpi},  when $r<A$, the
dominant contribution to the numerator in Eq.  \ref{Eq:momentpi} is
obtained from $x\sim 1/Q$,  and the moment $<b(Q)>_\pi\sim\ const$.
Due to the inherent unknowns of the wave function's dependence on
both $x$ and $b$, it is not possible to establish suppression of the
contributions of quark $OAM$ in pQCD in a model-independent way.

\begin{figure}[]
\centering
\includegraphics[]{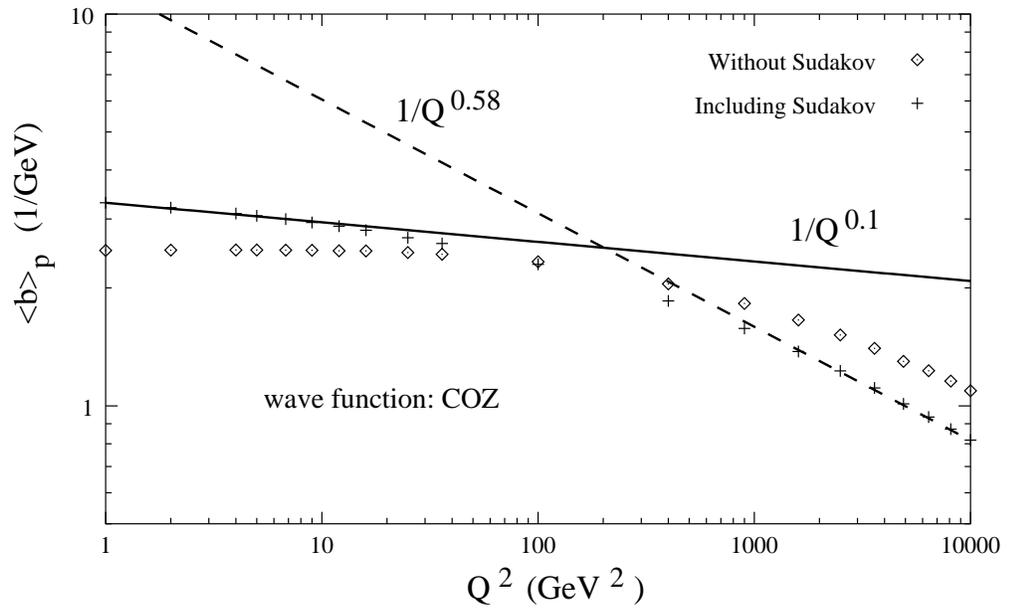}
\caption{The moment $<b_2>$ of the proton form factor kernel with and
without including the Sudakov form factor.  The COZ model is used for
the $x$-dependence of wave functions.  }
\label{fig:moment}
\end{figure}

Similar results are obtained for the proton.  In Fig.
\ref{fig:moment} we plot one moment of the transverse separation $b_2$
for the proton using the COZ model for the $x$-dependence of wave
functions \cite{COZ}, both including and not including the Sudakov
form factor \cite{LiSterman}.  It is clear that over a very large
range the moment has a very weak dependence on $Q$.  Hence the form
factor ratio $R(Q)= QF_2/F_1$ scales like a constant in this region,
in agreement with the recent JLAB experimental result.  Asymptotic
analysis \cite{jain03} shows that for a wide range of x-dependence in wave
functions, the form factor ratio $R(Q)$ scales like a constant, modulo
logarithms.

{\bf Acknowledgments}  Supported in part under DEO grant number
DE-FG03-98ER41079.

\begin{thebibliography}{99}

\bibitem{jones} M.K. Jones {\it et al},  Phys. Rev. Lett. {\bf 84},
       1398 (2000).
\bibitem{Gayou} O. Gayou {\it et al}, Phys.  Rev.  Lett.  {\bf{88}},
         092301, (2002).
\bibitem{Perdrisat} C. F. Perdrisat, these proceedings.

  \bibitem{BNL93} P. Jain and J. P. Ralston, in {\it Future Directions
in Particle and Nuclear Physics at Multi-GeV Hadron Beam Facilities}
(Proceedings of the Workshop held at BNL, 4-6 March, 1993),
      hep-ph/9305250.
\bibitem{Buniy} R. Buniy, J. P. Ralston, and P. Jain, in {\it VII
         International Conference on the Intersections of Particle and Nuclear
     Physics} (Quebec City, 2000) edited by Z. Parseh and W. Marciano (AIP,
         NY 2000), hep-ph/0206074.
\bibitem{Bologna}    J. P. Ralston, R. V. Buniy and P. Jain
	 {\it Proceedings of DIS 2001, 9th International Workshop on
	 Deep Inelastic Scattering}, Bologna, 27 April - 1 May, 2001,
	 hep-ph/0206063.
\bibitem{jain03} J. P. Ralston and P. Jain, hep-ph/0302043.
\bibitem{BL80} G. P. Lepage and S. J. Brodsky, Phys. Rev. D {\bf 22},
    2157 (1980).
\bibitem{BL} S. J. Brodsky and G. P. Lepage, Phys. Rev. {\bf D 24},
      2848 (1981).
\bibitem{RP}B.~Pire and J.~P.~Ralston, Phys.\ Lett.\ B {\bf 117}, 233 (1982).
     \bibitem{BFMatveev} S. J. Brodsky and G. R. Farrar, Phys. Rev. {\bf D 11},
        1309 (1975); V. A. Matveev, R. M. Muradian and A. N. Tavkhelidze,
     Lett. Nuovo Cim. {\bf 7}, 719 (1973).
\bibitem{Li} H.-N. Li Phys. Rev.  {\bf D 48},
    4243 (1993).
\bibitem{BJY} A. V. Belitsky, X. Ji and F. Yuan,
hep-ph/0212351.
\bibitem{miller} M. R. Frank, B. K. Jennings and G.A. Miller,
Phys. Rev. {\bf {C54}}, 920 (1996); G.A. Miller and M. R. Frank,
nucl-th/0201021.
\bibitem{LiSterman}
H.-N. Li and G.
Sterman, Nucl. Phys. {\bf B 381}, 129 (1992).
\bibitem{COZ}
V. L. Chernyak, A. A. Ogloblin, I. R. Zhitnitsky
  Sov. J. Nucl. Phys. {\bf 48}, 536 (1988); Yad. Fiz. {\bf 48},
      841 (1988).
\end{thebibliography}
\end{document}